# A New Estimator For Population Mean Using Two Auxiliary Variables in Stratified random Sampling

Rajesh Singh and Sachin Malik

Department of Statistics, Banaras Hindu University

Varanasi-221005, India

(sachinkurava999@gmail.com, rsinghstat@gmail.com)

**Abstract**

In this paper, we suggest an estimator using two auxiliary variables in stratified random sampling. The propose estimator has an improvement over mean per unit estimator as well as some other considered estimators. Expressions for bias and MSE of the estimator are derived up to first degree of approximation. Moreover, these theoretical findings are supported by a numerical example with original data.

## 1. Introduction

The problem of estimating the population mean in the presence of an auxiliary variable has been widely discussed in finite population sampling literature. Out of many ratio, product and regression methods of estimation are good examples in this context. Diana [2] suggested a class of estimators of the population mean using one auxiliary variable in the stratified random sampling and examined the MSE of the estimators up to the $k^{th}$ order of approximation. Kadilar and Cingi [3], Singh et al. [7], Singh and Vishwakarma [8], Koyuncu and Kadilar [4] proposed estimators in stratified random sampling. Singh [9] and Perri [6] suggested some ratio cum product estimators in simple random sampling. Bahl and Tuteja [1] and Singh et al. [11] suggested some exponential ratio type estimators. In this chapter, we suggest some exponential-type estimators using the auxiliary information in the stratified random sampling.

Consider a finite population of size N and is divided into L strata such that $\sum_{h=1}^{L} N_h = N$ where $N_h$ is the size of $h^{th}$ stratum (h=1,2,...,L). We select a sample of size $n_h$ from each stratum by simple random sample without replacement sampling such that $\sum_{h=1}^{L} n_h = n$, where $n_h$ is the stratum sample size. A simple random sample of size $n_h$ is drawn without replacement from the $h_{th}$ stratum such that $\sum_{h=1}^{L} n_h = n$. Let ($y_{hi}$, $x_{hi}$, $z_{hi}$) denote the observed values of y, x, and z on the $i_{th}$ unit of the $h_{th}$ stratum, where i=1, 2, 3...$N_h$.

To obtain the bias and MSE, we write

$$\bar{y}_{st} = \sum_{h=1}^{L} w_h \bar{y}_h = \bar{Y}(1+e_0), \quad \bar{x}_{st} = \sum_{h=1}^{L} w_h \bar{x}_h = \bar{X}(1+e_1), \quad \bar{z}_{st} = \sum_{h=1}^{L} w_h \bar{z}_h = \bar{Z}(1+e_2)$$

Such that,

$$E(e_0) = E(e_0) = E(e_0) = 0$$

$$V_{rst} = \sum_{h=1}^{L} w_h^{r+s+t} \frac{E\left[(\bar{y}_h - \bar{Y})^r (\bar{x}_h - \bar{X})^s (\bar{z}_h - \bar{Z})^t\right]}{\bar{Y}^r \bar{X}^s \bar{Z}^t}$$

where,

$$\bar{y}_{st} = \sum_{h=1}^{L} w_h \bar{y}_h, \quad \bar{y}_h = \frac{1}{n_h} \sum_{i=1}^{n_h} \bar{y}_{hi}, \quad \bar{Y}_h = \frac{1}{N_h} \sum_{i=1}^{n_h} \bar{Y}_{hi}$$

$$Y = \bar{Y}_{st} = \sum_{h=1}^{L} w_h \bar{Y}_h, \quad w_h = \frac{N_h}{N}$$

and

$$V(\bar{y}_{st}) = \bar{Y}^2 V_{200} \tag{1.1}$$

Similar expressions for X and Z can also be defined.

$$E\left(e_0^2\right)=\frac{\sum_{h=1}^{L}W_h^2 f_h S_{yh}^2}{\overline{Y}^2}=V_{200}, \qquad E\left(e_1^2\right)=\frac{\sum_{h=1}^{L}W_h^2 f_h S_{xh}^2}{\overline{X}^2}=V_{020},$$

$$E\left(e_2^2\right)=\frac{\sum_{h=1}^{L}W_h^2 f_h S_{zh}^2}{\overline{Z}^2}=V_{002}, \qquad E\left(e_0 e_1\right)=\frac{\sum_{h=1}^{L}W_h^2 f_h S_{yxh}^2}{\overline{Y}\overline{X}}=V_{110},$$

$$E\left(e_0 e_2\right)=\frac{\sum_{h=1}^{L}W_h^2 f_h S_{yzh}^2}{\overline{Y}\overline{Z}}=V_{101}, \qquad \text{and} E\left(e_1 e_2\right)=\frac{\sum_{h=1}^{L}W_h^2 f_h S_{xzh}^2}{\overline{X}\overline{Z}}=V_{011},$$

where , $S_{yh}^2=\sum_{i=1}^{N_h}\frac{\left(\overline{y_h}-\overline{Y}_h\right)^2}{N_h-1}$, $\quad S_{xh}^2=\sum_{i=1}^{N_h}\frac{\left(\overline{x}_h-\overline{X}_h\right)^2}{N_h-1}$

$S_{zh}^2=\sum_{i=1}^{N_h}\frac{\left(\overline{z}_h-\overline{Z}_h\right)^2}{N_h-1}$, $\quad S_{yzh}=\sum_{i=1}^{N_h}\frac{\left(\overline{z}_h-\overline{Z}_h\right)\left(\overline{y_h}-\overline{Y}_h\right)}{N_h-1}$, $\quad S_{xzh}=\sum_{i=1}^{N_h}\frac{\left(\overline{x}_h-\overline{X}_h\right)\left(\overline{z_h}-\overline{Z}_h\right)}{N_h-1}$

and $f_h=\frac{1}{n_h}-\frac{1}{N_h}$

## 2. **Estimators in literature**

In order to have an estimate of the study variable y, assuming the knowledge of the population proportion P, Naik and Gupta [5] and Singh et al. [11] respectively proposed following estimators

$$t_1=\overline{y}_{st}\left(\frac{\overline{X}}{\overline{x}_{st}}\right) \tag{2.1}$$

$$t_2=\overline{y}_{st}\exp\left(\frac{\overline{X}-\overline{x}_{st}}{\overline{X}+\overline{x}_{st}}\right) \tag{2.2}$$

The MSE expressions of these estimators are given as

$$MSE(t_1) = \bar{Y}^2 \left[V_{200} + V_{020} - 2V_{110}\right] \quad (2.3)$$

$$MSE(t_2) = \bar{Y}^2 \left[V_{200} + \frac{V_{020}}{4} - V_{110}\right] \quad (2.4)$$

When the information on the two auxiliary variables is known, Singh [10] proposed some ratio cum product estimators in simple random sampling to estimate the population mean of the study variable y.

Motivated by Singh [10] and Singh et al. [7], Singh and kumar propose some estimators in stratified sampling as

$$t_3 = \bar{y}_{st} \exp\left[\frac{\bar{X} - \bar{x}_{st}}{\bar{X} + \bar{x}_{st}}\right] \exp\left[\frac{\bar{Z} - \bar{z}_{st}}{\bar{Z} + \bar{z}_{st}}\right] \quad (2.5)$$

$$t_4 = \bar{y}_{st} \exp\left[\frac{\bar{x}_{st} - \bar{X}}{\bar{x}_{st} + \bar{X}}\right] \exp\left[\frac{\bar{z}_{st} - \bar{Z}}{\bar{z}_{st} + \bar{Z}}\right] \quad (2.6)$$

$$t_5 = \bar{y}_{st} \exp\left[\frac{\bar{X} - \bar{x}_{st}}{\bar{X} + \bar{x}_{st}}\right] \exp\left[\frac{\bar{z}_{st} - \bar{Z}}{\bar{z}_{st} + \bar{Z}}\right] \quad (2.7)$$

$$t_6 = \bar{y}_{st} \exp\left[\frac{\bar{x}_{st} - \bar{X}}{\bar{x}_{st} + \bar{X}}\right] \exp\left[\frac{\bar{Z} - \bar{z}_{st}}{\bar{Z} + \bar{z}_{st}}\right] \quad (2.8)$$

The MSE equations of these estimators can be written as

$$MSE(t_3) = \bar{Y}^2 \left[V_{200} + \frac{V_{020}}{4} + \frac{V_{002}}{4} - V_{110} - V_{101} + \frac{V_{011}}{2}\right] \quad (2.9)$$

$$MSE(t_4) = \bar{Y}^2 \left[V_{200} + \frac{V_{020}}{4} + \frac{V_{002}}{4} + V_{110} + V_{101} + \frac{V_{011}}{2}\right] \quad (2.10)$$

$$MSE(t_5) = \bar{Y}^2 \left[V_{200} + \frac{V_{020}}{4} + \frac{V_{002}}{4} - V_{110} + V_{101} - \frac{V_{011}}{2}\right] \quad (2.11)$$

$$MSE(t_6) = \overline{Y}^2\left[V_{200} + \frac{V_{020}}{4} + \frac{V_{002}}{4} + V_{110} - V_{101} - \frac{V_{011}}{2}\right] \tag{2.12}$$

When there are two auxiliary variables, the regression estimator of $\overline{Y}$ will be

$$t_7 = \overline{y}_{st} + b_{1h}\left(\overline{X} - \overline{x}_{st}\right) + b_{2h}\left(\overline{Z} - \overline{z}_{st}\right) \tag{2.13}$$

Where $b_{1h} = \frac{s_{yx}}{s_x^2}$ and $b_{2h} = \frac{s_{yz}}{s_z^2}$. Here $s_x^2$ and $s_z^2$ are the sample variances of x and z respectively, $s_{yx}$ and $s_{yz}$ are the sample covariance's between y and x and between z respectively. The MSE expression of this estimator is:

$$MSE\left(t_7\right) = \sum_{h=1}^{L} W_h^2 f_h S_{yh}^2 \left(1 - \rho_{yxh}^2 - \rho_{yzh}^2 + 2\rho_{yxh}\rho_{yzh}\rho_{xzh}\right) \tag{2.14}$$

**3. The proposed estimator**

We suggest using the ratio estimator given in equation (2.5) instead of estimator given in equation (2.13). By this way, we obtain the following estimator

$$t_p = \overline{y}_{st}\exp\left[\frac{\overline{X} - \overline{x}_{st}}{\overline{X} + \overline{x}_{st}}\right]^{m1} \exp\left[\frac{\overline{Z} - \overline{z}_{st}}{\overline{Z} + \overline{z}_{st}}\right]^{m2} + b_{1h}\left(\overline{X} - \overline{x}_{st}\right) + b_{2h}\left(\overline{Z} - \overline{z}_{st}\right) \tag{3.1}$$

Expressing equation (3.1) in terms of e's, we have

$$t_p = \overline{Y}\left(1 + e_0\right)\left\{\exp\left(\frac{-e_1}{2 + e_1}\right)^{m_1} \exp\left(\frac{-e_2}{2 + e_2}\right)^{m_2}\right\} - b_{1h}\overline{X}e_1 - b_{2h}\overline{Z}e_2$$

$$= \overline{Y}\left[1 + e_0 - \frac{m_1 e_1}{2} + \frac{m_1^2 e_1^2}{4} - \frac{m_2 e_2}{2} - \frac{m_1 m_2 e_1 e_2}{4} + \frac{m_2^2 e_2^2}{4} - \frac{m_2 e_0 e_2}{2} - \frac{m_1 e_0 e_1}{2}\right]$$

$$- b_{1h}e_1\overline{X} - b_{2h}e_2\overline{Z} \tag{3.2}$$

Squaring both sides of (3.2) and neglecting the term having power greater than two, we have

$$\left(t_p - \overline{Y}\right)^2 = \left\{\overline{Y}\left[e_0 - \frac{m_1 e_1}{2} - \frac{m_2 e_2}{2}\right] - b_{1h} e_1 \overline{X} - b_{2h} e_2 \overline{Z}\right\}^2 \tag{3.3}$$

Taking expectations of both the sides of (3.3), we have the mean squared error of $t_p$ up to the first degree of approximation as

$$MSE(t_p) = \overline{Y}^2\left[V_{200} + P_1\right] + P_2 - \overline{Y}P_3 \tag{3.4}$$

Where,

$$\left.\begin{aligned} P_1 &= \frac{m_1^2 V_{020}}{4} + \frac{m_2^2 V_{002}}{4} + \frac{m_1 m_2 V_{011}}{2} - m_1 V_{110} - m_2 V_{101} \\ P_2 &= B_{1h}^2 V_{020} + B_{2h}^2 V_{002} + 2B_{1h} B_{2h} V_{011} \\ P_3 &= -2B_{1h} V_{110} - 2B_{2h} V_{101} + m_1 B_{1h} V_{020} + m_1 B_{2h} V_{011} + m_2 B_{1h} V_{011} + m_2 B_{2h} V_{002} \end{aligned}\right\} \tag{3.5}$$

Where, $B_{1h} = \dfrac{\sum_{h=1}^{L} W_h^2 f_h \rho_{yxh} S_{yh} S_{xh}}{\sum_{h=1}^{L} W_h^2 f_h S_{xh}^2}$ and $B_{2h} = \dfrac{\sum_{h=1}^{L} W_h^2 f_h \rho_{yzh} S_{yh} S_{zh}}{\sum_{h=1}^{L} W_h^2 f_h S_{zh}^2}$

The optimum values of $m_1$ and $m_2$ will be

$$\left.\begin{aligned} m_1 &= \frac{4\left[B_{1h} V_{011} V_{002} + B_{2h} V_{011}^2 - B_{1h} V_{020} V_{002} - B_{2h} V_{011} V_{002}\right]}{\overline{Y}\left[V_{020} V_{002} - V_{011}^2\right]} \\ m_2 &= \frac{4\left[B_{1h} V_{011} V_{020} + B_{2h} V_{011}^2 - B_{1h} V_{011} V_{020} - B_{2h} V_{002} V_{020}\right]}{\overline{Y}\left[V_{020} V_{002} - V_{011}^2\right]} \end{aligned}\right\} \tag{3.6}$$

Putting optimum values of $m_1$ and $m_2$ from (3.6), we obtained min MSE of proposed estimator $t_p$.

## 4. Efficiency comparison

In this section, the conditions for which the proposed estimator $t_p$ is better than $\overline{y}_{st}$, $t_1$, $t_2$, $t_3$,

$t_4$, $t_5$, $t_6$, and $t_7$.

The variance is given by

$$V(\bar{y}_{st}) = \bar{Y}^2 V_{200} \tag{4.1}$$

To compare the efficiency of the proposed estimator with the existing estimator, from (4.1) and (2.3), (2.4), (2.9), (2.10), (2.11), (2.12) and (2.14), we have

$$V(\bar{y}_{st}) - MSE(t_p) = \bar{Y}^2 P_1 + P_2 - \bar{Y}P_3 \geq 0 \tag{4.2}$$

$$MSE(t_1) - MSE(t_p) = \bar{Y}^2 [V_{020} - 2V_{110}] - \bar{Y}^2 P_1 - P_2 + \bar{Y}P_3 \geq 0 \tag{4.3}$$

$$MSE(t_2) - MSE(t_p) = \bar{Y}^2 \left[\frac{V_{020}}{4} - V_{110}\right] - \bar{Y}^2 P_1 - P_2 + \bar{Y}P_3 \geq 0 \tag{4.4}$$

$$MSE(t_3) - MSE(t_p) = \bar{Y}^2 \left[\frac{V_{020}}{4} + \frac{V_{002}}{4} - V_{110} - V_{101} + \frac{V_{011}}{2}\right] - \bar{Y}^2 P_1 - P_2 + \bar{Y}P_3 \geq 0 \tag{4.5}$$

$$MSE(t_4) - MSE(t_p) = \bar{Y}^2 \left[\frac{V_{020}}{4} + \frac{V_{002}}{4} + V_{110} + V_{101} + \frac{V_{011}}{2}\right] - \bar{Y}^2 P_1 - P_2 + \bar{Y}P_3 \geq 0 \tag{4.6}$$

$$MSE(t_5) - MSE(t_p) = \bar{Y}^2 \left[\frac{V_{020}}{4} + \frac{V_{002}}{4} - V_{110} + V_{101} + \frac{V_{011}}{2}\right] - \bar{Y}^2 P_1 - P_2 + \bar{Y}P_3 \geq 0 \tag{4.7}$$

$$MSE(t_6) - MSE(t_p) = \bar{Y}^2 \left[\frac{V_{020}}{4} + \frac{V_{002}}{4} + V_{110} - V_{101} + \frac{V_{011}}{2}\right] - \bar{Y}^2 P_1 - P_2 + \bar{Y}P_3 \geq 0 \tag{4.8}$$

Using (4.2) - (4.8), we conclude that the proposed estimator outperforms than the estimators considered in literature.

## 5. Empirical study

In this section, we use the data set in Koyuncu and Kadilar [4]. The population statistics are given in Table 3.2.1. In this data set, the study variable (Y) is the number of teachers, the first auxiliary variable (X) is the number of students, and the second auxiliary variable (Z) is the number of classes in both primary and secondary schools.

**Table 5.1: Data statistics**

| | | |
|---|---|---|
| $N_1=127$ | $N_2=117$ | $N_3=103$ |
| $N_4=170$ | $N_5=205$ | $N_6=201$ |
| $n_1=31$ | $n_2=21$ | $n_3=29$ |
| $n_4=38$ | $n_5=22$ | $n_6=39$ |
| $S_{y1}=883.835$ | $S_{y2}=644$ | $S_{y3}=1033.467$ |
| $S_{y4}=810.585$ | $S_{y5}=403.654$ | $S_{y6}=711.723$ |
| $\overline{Y}_1=703.74$ | $\overline{Y}_2=413$ | $\overline{Y}_3=573.17$ |
| $\overline{Y}_4=424.66$ | $\overline{Y}_5=267.03$ | $\overline{Y}_6=393.84$ |
| $S_{x1}=30486.751$ | $S_{x2}=15180.760$ | $S_{x3}=27549.697$ |
| $S_{x4}=18218.931$ | $S_{x5}=8997.776$ | $S_{x6}=23094.141$ |
| $\overline{X}_1=20804.59$ | $\overline{X}_2=9211.79$ | $\overline{X}_3=14309.30$ |
| $\overline{X}_4=9478.85$ | $\overline{X}_5=5569.95$ | $\overline{X}_6=12997.59$ |
| $S_{yx1}=25237153.52$ | $S_{yx2}=9747942.85$ | $S_{yx3}=28294397.04$ |
| $S_{yx4}=1452885.53$ | $S_{yx5}=3393591.75$ | $S_{yx6}=15864573.97$ |
| $\rho_{yx1}=0.936$ | $\rho_{yx2}=0.996$ | $\rho_{yx3}=0.994$ |
| $\rho_{yx4}=0.983$ | $\rho_{yx5}=0.989$ | $\rho_{yx6}=0.965$ |
| $\beta_2(x_1)=4.593$ | $\beta_2(x_2)=18.543$ | $\beta_2(x_3)=15.446$ |
| $\beta_2(x_4)=10.162$ | $\beta_2(x_5)=21.947$ | $\beta_2(x_6)=23.114$ |

$\beta_2(y_1) = 2.158$ $\beta_2(y_2) = 16.392$ $\beta_2(y_3) = 14.979$

$\beta_2(y_4) = 12.167$ $\beta_2(y_5) = 21.008$ $\beta_2(y_6) = 20.254$

$S_{z1} = 555.5816$ $S_{z2} = 365.4576$ $S_{z3} = 612.9509$

$S_{z4} = 458.0282$ $S_{z5} = 260.8511$ $S_{z6} = 397.0481$

$\bar{Z}_1 = 498.28$ $\bar{Z}_2 = 318.33$ $\bar{Z}_3 = 431.36$

$\bar{Z}_4 = 498.28$ $\bar{Z}_5 = 227.20$ $\bar{Z}_6 = 313.71$

$S_{yz1} = 480688.2$ $S_{yz2} = 230092.8$ $S_{yz3} = 623019.3$

$S_{yz4} = 36493.4$ $S_{yz5} = 101539$ $S_{yz6} = 277696.1$

$S_{xz1} = 15914648$ $S_{xz2} = 5379190$ $S_{xz3} = 16490067456$

$S_{xz4} = 8041254$ $S_{xz5} = 214457$ $S_{xz6} = 8857729$

$\rho_{yz1} = 0.978$ $\rho_{yz2} = 0.976$ $\rho_{yz3} = 0.983$

$\rho_{yz4} = 0.982$ $\rho_{yz5} = 0.964$ $\rho_{yz6} = 0.982$

$\beta_2(z_1) = 2.314$ $\beta_2(z_2) = 11.190$ $\beta_2(z_3) = 10.786$

$\beta_2(z_4) = 8.624$ $\beta_2(z_5) = 9.720$ $\beta_2(z_6) = 14.406$

We have computed the pre relative efficiency (PRE) of different estimators of $\bar{Y}_{st}$ with respect to $\bar{y}_{st}$ and complied in table 5.2:

**Table 5.2: Percent Relative Efficiencies (PRE) of estimator**

| S. No. | Estimators | PRE'S |
|---|---|---|
| 1 | $\bar{y}_{st}$ | 100 |
| 2 | $t_1$ | 1029.46 |
| 3 | $t_2$ | 370.17 |
| 4 | $t_3$ | 2045.43 |
| 5 | $t_4$ | 27.94 |
| 6 | $t_5$ | 126.41 |
| 7 | $t_6$ | 77.21 |
| 8 | $t_7$ | 2360.54 |
| 9 | $t_p$ | 4656.35 |

## 6. Conclusion

In this paper, we proposed a new estimator for estimating unknown population mean of study variable using information on two auxiliary variables. Expressions for bias and MSE of the estimator are derived up to first degree of approximation. The proposed estimator is compared with usual mean estimator and other considered estimators. A numerical study is carried out to support the theoretical results. In the table 5.2, the proposed estimator performs better than the usual sample mean and other considered estimators.